\documentclass{aastex}
\usepackage{spr-astr-addons}
\usepackage{url}

\usepackage{aeguill}

\RequirePackage{color}

\begin{document}

\title{MOSAIX: a tool to built large mosaics from GALEX images}
\slugcomment{Not to appear in Nonlearned J., 45.}
\shorttitle{MOSAIX}
\shortauthors{Armengot et al.}

\author{Marcelo Armengot} \and \author{Nestor S\'anchez} \and \author{Javier L\'opez-Santiago} \and \author{Ana In\'es G\'omez de Castro}
\email{aig@ucm.es}

\altaffiltext{}{AEGORA Research Group - Universidad Complutense de Madrid, 
Fac. de CC Matem\'aticas, 28040 Madrid, Spain}

\begin{abstract}
Large sky surveys are providing a huge amount of information for studies of the
interstellar medium, the galactic structure or the cosmic web. Setting into a common
frame information coming from different wavelengths, over large fields of view, is
needed for this kind of research. GALEX is the only nearly
all-sky survey at ultraviolet wavelengths and contains fundamental information
for all types of studies. GALEX field of view is circular embedded
in a squared matrix of 3840 $\times$ 3840 pixels. This fact makes it hard to get GALEX 
images properly overlapped with the existing astronomical tools such as Aladin or Montage. 
We developed our own software for this purpose. In this article, we describe this 
software and makes it available to the community.

\end{abstract}




\section{Introduction}
\label{intro}

Mosaicking astronomical images is a complex task. Sky images are projections of spherical 
maps onto the Euclidean plane. Each astronomical image may have its own projection system 
(the world Coordinate System standard proposes up to 25 different projections). Thus, 
combining sky images into a unique image (a mosaic) can involve not only rotating and translating, 
but also re-projecting the images. 

Images from astronomical facilities are distributed in Flexible Image Transport System
(FITS) file format. FITS is a digital file format used to store scientific data:
images, binary tables and, in general, data arrays of arbitrary dimension. Each FITS file 
consists of a header and a data block: a table, an image, a spectrum, a list of photons or a data cube. 
The header is an ASCII text that contains the information describing the data set, the instrument and
its configuration to guarantee that the observation can be repeated by any other observer,
according to the requirements of the scientific method (see, for instance the documentation in the
International Virtual Observatory Alliance for astronomical data standardizing at \texttt{www.ivoa.net}).
In particular, the header of 2D- or 3D-images offers information about the scientific coordinate systems that are overlaid 
on the image itself (typically only one) including the projection system. FITS visualization tools 
convert from pixels to astronomical coordinates by using this information. Therefore, the header 
contains the information needed to combine and re-project images onto mosaics.

\begin{figure}[!h]
	\centering
	\includegraphics[trim=0mm 0mm 0mm -5mm,width=7cm]{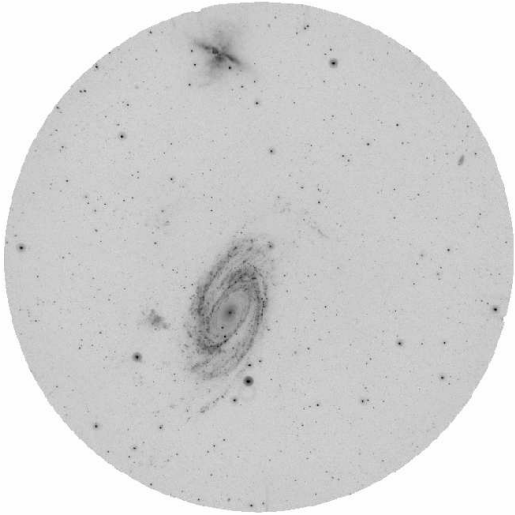} 
	\caption{GALEX NUV image of the M81 and M82 galaxies with the field of view overlaid.}
	\label{fig1}
\end{figure} 

However, building a mosaic involves more information than just pure astrometry. Even in well planned
space-based surveys, such as the all-sky survey runs by the Galactic Evolution Explorer (GALEX-AIS), exposure
times are not exactly the same, nor background levels. Thus, building a scientifically useful
mosaic requires to correct for background levels, as well as to define algorithms that include 
flux rebinning when geometric re-projection is applied.

Aladin is an interactive software developed and maintained
by the Strasbourg astronomical Data Center (CDS) and used
extensively by the astronomical community to visualize
images coming from different sources and surveys. The astronomical image mosaic 
engine developed at Caltech/JPL for its space missions 
is Montage\footnote{Visit \texttt{http://montage.ipac.caltech.edu} for more details.}, a
toolkit for assembling FITS images into custom mosaics. However, it does not suit well 
for mosaicking GALEX images. The field of view of GALEX is circular (see Figure~\ref{fig1}). 
The FITS file contains zeros outside the field of view. The Montage task \emph{mFixNan} can 
be used to convert a range of supplied values into NaNs. One may be tempted to use 
this task to convert every pixel with a value equal or lower than zero into a NaN. 
However, this approach causes several problems when the resulting images with NaN 
values are mosaicked with Montage. The exposed field of the image may also contain 
pixels with a value of zero and they would not be treated by Montage.

\begin{figure}[!h]
	\centering
	\includegraphics[clip=true,trim=10mm 10mm 10mm 30mm,width=6cm]{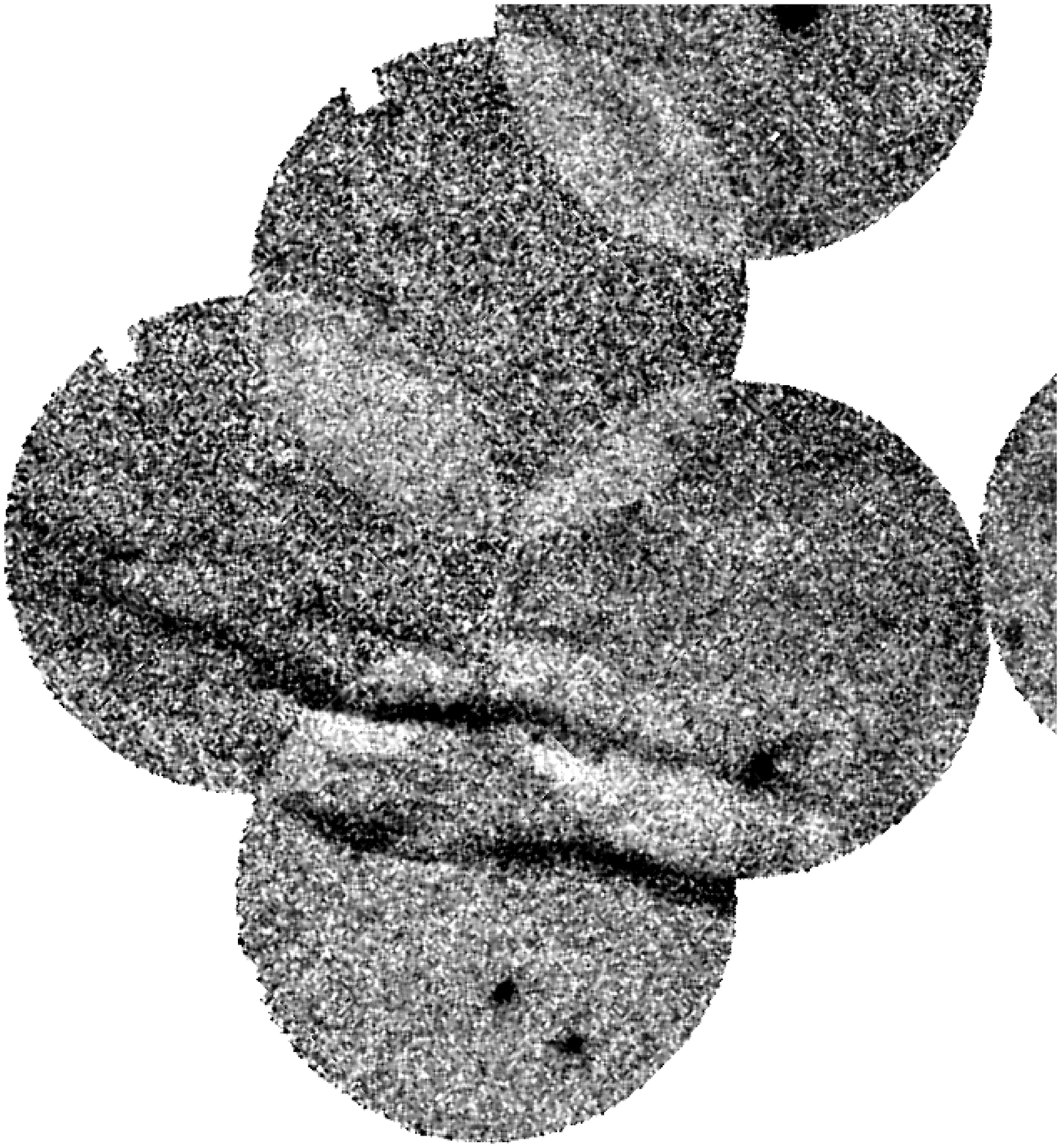} 
	\caption{GALEX FUV image of an Interstellar Medium filament 
          covering more than 2 degrees in the sky. This 
		       mosaic was generated with Aladin.}
	\label{fig2}
\end{figure}

The next UV mission to be flown is the World Space Observatory - Ultraviolet (WSO-UV) \citep{shustov3}. 
WSO-UV is a 170~cm primary telescope equipped with instrumentation for
imaging and spectroscopy in the ultraviolet (UV) range, from 115 to 320~nm. ISSIS is the imaging 
instrument on board WSO-UV and it will be equipped with Micro-Channel-Plate (MCP) detectors, as the
GALEX mission, which have very low read-out noise (to detect weak extended structures) but that can
be damaged by high count rates \citep{issis0, issis}. The ISSIS team will extensively use 
the GALEX database to select interstellar regions for future research and to define  avoidance regions. 
For this purpose, we require large UV maps of the sky where interstellar features extending 
over several degrees in the sky (see Figure ~\ref{fig2}) can be properly studied. Given the 
limitations of the Montage software, we developed our own software, MOSAIX, that 
we make available (subsection \ref{download}) to the scientific community.

The paper is organized as follows. In Section \ref{previa} a brief outline on GALEX data is provided. 
The algorithm is described in Section \ref{algorithm}  and the tests we have run are detailed 
in Section \ref{test}. 
Further useful information and implementation details can be found in Section \ref{apendice2}. 
Conclusions and future work are summarized in Section~\ref{epilogue}.

\section{About GALEX images}

\label{previa}


GALEX performed an all-sky survey in the far ultraviolet (FUV) and near ultraviolet (NUV) bands
\citep{galex1}. Releases from the mission include the archive stored in the 
NASA Mikulski Archive for Space Telescopes (MAST) as well as high level science products like
the point source catalogue \citep{bianchi}. Our work uses images
obtained by GALEX in the far ultraviolet, 135-175 nm, or FUV band and in the 
near ultraviolet, 175-280 nm, or NUV band. 
The current release of the GALEX archive contains data products 
from the mission pipeline\footnote{Visit the general FAQ
of GR6 in \texttt{galex.stsci.edu} for a complete list of products.}. Most of them 
are FITS files, both tables and images. 
The archive contains flux-calibrated images in the NUV and FUV bands,
in photons per second per pixel, as well as background-substracted intensity maps and 
count maps (photons/pixel) in both NUV and FUV. For this work, intensity maps without 
background subtraction were used because it is a better choice when looking for extended
weak structures. However, for high sensitivity applications we recommend the user to
substract the sky background prior to mosaicking. Note that the GALEX mission provides
background substracted images among its standard output products.




The information contained in the FITS header is used to compute the projection elements for building the mosaic.
	NAXIS1 and NAXIS2 are the sizes (in pixels) of the data for the horizontal and vertical axis,
	respectively. The GALEX mission database provides $3840 \times 3840$ pixels images. The RA\_CENT and
	DEC\_CENT values are the right ascension and declination of the target point and they correspond to
	the central pixel coordinates ($1920.5,1920.5$) that are given by the parameters CRPIX1 and CRPIX2.
	The increase step for each axis is given by CDELT1 and CDELT2 and it is always $=4.1667 \times 10^{-4}$
	degrees/pixel (equivalent to 1.6 degrees / 3840 pixels). The angle between the North and the second 
        axis (CROTA2) is always zero for the
	GALEX images we are using in this work. With this information read from the header the coordinates
	of every point on the image can be computed. The matrix of pixels representing the image has only one
	channel with a given view size of $1.28$ degrees (in FUV) or $1.24$ degrees (NUV).

\section{Merging algorithm}

\label{algorithm}

GALEX images are generated in gnomonic projection \citep{galex1}, i.e. the
celestial sphere is projected on the plane of the sky considered to be tangent to
the sphere in the center of the field \citep{green}. Projection effects are not 
considered at this stage of the software development. Distortion effects are 
negligible\footnote{The difference in pixel sizes between the center and the edge
of the image is $\sim 0.2$ mas.} for the GALEX field of view. For the correct
image alignment both the angular distance and the rotation angle between
the images must be known. We first present how two images are 
assembled together (Section \ref{caso1}) and then we generalize it (Section \ref{caso2}) 
for more than two images. 

\subsection{Two images}

        \label{caso1}
The alignment of the reference system is made in two steps: displacement
(\ref{ecuaciones_desplazamiento}) to match the center of the two images and then, rotation 
(\ref{ecuaciones_rotacion}). The generic procedure can be summarized as follows:
First, we build  the required canvas (in a matrix grid) for the mosaic and then
the first image is mapped to it without any motion or 
rotation. We name this first image the \emph{baseline image}. Finally,
the \emph{inset image} is added to the canvas after moving
and rotating it to completely match the baseline image in overlapping portions.

	\subsubsection{Translation}	

        \label{ecuaciones_desplazamiento}

	Let ($C_x$,$C_y$) be the center of the baseline image in the 
        new canvas and, associated to this reference pixel, the coordinates of the
        baseline image must be read from the header. 
        Let ($\alpha_0$,$\delta_0$) $\in \mathbb{R}$ be the 
        coordinates of this point which are the right ascension ($\alpha_0$)
        and declination ($\delta_0$) respectively. Let ($\alpha,\delta$) $\in \mathbb{R}$ be 
        the center coordinates of the inset image.

        The distances $\Delta x$ and $\Delta y$ (in pixels) between the center of the two images  
        are computed \citep{gnomonic} as,
	\begin{eqnarray}	
	    A      & = & \cos(\delta)   \times   \cos(\alpha-\alpha_{0})  \label{eqmv1}  \\
	    F      & = & s   \times   
            \frac{\left(180/\pi \right)}{\sin(\delta_{0})  
              \times  \sin(\delta)} + (A  \times  \cos(\delta_{0})) \label{eqmv2} \\
	    \Delta y & = & -F  \times  (\cos(\delta_{0})  \times  \sin(\delta)) - (A  \times  \sin(\delta_{0})) \label{eqmv3} \\
            \Delta x &  = & -F  \times  \cos(\delta) \times  \sin(\alpha-\alpha_{0}) \label{eqmv4} 
	\end{eqnarray}	
        being $s$ the pixel scale. The variables $\Delta x$ and $\Delta y$ fit the offset 
        between the two images but they must be integers, thus we use the $round$ function to 
        find the nearest integer:
	\begin{eqnarray}	
				X & = & round (C_{x} + \Delta x) \nonumber \\
				Y & = & round (C_{y} + \Delta y)
	\end{eqnarray}

\subsubsection{Rotation}

        \label{ecuaciones_rotacion}

        In Figure~\ref{angle} the orientation of the images in the plane of the sky is shown. 
        Both the center of the baseline image ($\alpha_0,\delta_0$) and the center 
        of the inset image ($\alpha,\delta$) are marked.


        If the position angle in the FITS file header is zero
no rotation needs to be applied and the y-axis of the digital image, the vertical line 
in the figure, is aligned with the polar direction. As a result, the angle between the 
two images can be computed as:

     \begin{figure}[!ht]
	\centering
	\includegraphics[trim=0mm 0mm 0mm -5mm,width=6cm]{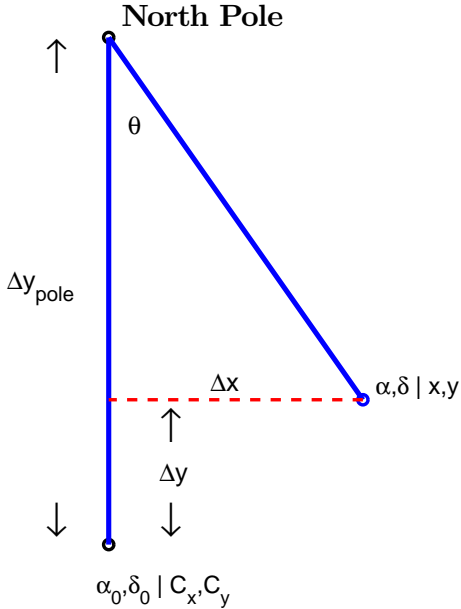}
	\caption{Sketch of the relative position of two GALEX images projected on the sky tangent plane.}
	\label{angle}
     \end{figure}

	\begin{eqnarray}
          \label{rot_equ}
                        \theta & = & \arctan \left( \frac{\Delta x}{\Delta y_{pole}-\Delta y} \right) 			
	\end{eqnarray}	


\subsection{More than two images}

	\label{caso2}
        When only two images are processed the coordinates of one of them
        can be used as a reference for the output mosaic. In multiple
        merging, the point located at the middle of the whole set of images
        is associated with the center pixel of the new FITS 
        image canvas. This point will be the baseline point and all the merging
        images will be shifted and rotated around it for being inserted into the canvas.

\subsubsection{Computing the canvas size}

From a set of FITS images with central coordinates ($\alpha_i,\delta_i$), 
the resulting canvas size is computed in four steps:

\begin{enumerate}
\item The first size estimate (in degrees) is computed from the centers
      of the images:
        \begin{eqnarray}
                \Delta \alpha & = & \max(\alpha_i) - \min(\alpha_i)  \nonumber \\
                \Delta \delta & = & \max(\delta_i) - \min(\delta_i) 
                \label{region}
        \end{eqnarray}

\item The pixel scale ($s$) is applied to convert angular distances into pixels.

\item Then, a frame is added to the canvas. If each single image has
      $N$ pixels per side, the canvas size ($W \times H$) will be:      
        \begin{eqnarray}
                H & = & N + ( \Delta \alpha \times s ) \nonumber \\
                W & = & N + ( \Delta \delta \times s )
                \label{canvas}
        \end{eqnarray}

\item Finally, the canvas size is made a little bit larger because
rotation effects may prevent that an image is properly inserted.
Note that the worse case is an image to be placed in the corner of the canvas 
that must be rotated by 45 degrees.  A pixel frame of around 795 pixels would be 
needed in this case.

\end{enumerate}

We must point out that the possibility to perform pixel resampling is not
offered in this first version of the software. That is, the final canvas pixel
size is the same as the original GALEX images (1.5 arcsec/pixel).

\subsubsection{Center (baseline) point}

        In a multiple image merging task, nearly all the pictures
        will be rotated. Our approach fixes a central reference
        point in the middle of the image with its space coordinates rightly defined 
        ($\alpha_c,\delta_c$). This is the midpoint between maximum and minimum 
        coordinates and it is associated to the central pixel ($W/2$,$H/2$) of the canvas:
        \begin{eqnarray}
		\alpha_c & = & \min(\alpha_i) + \frac{\Delta\alpha}{2} \nonumber \\
		\delta_c & = & \min(\delta_i) + \frac{\Delta\delta}{2}
                \label{centro_equ}
        \end{eqnarray}

\subsubsection{Algorithm}
        
        In Section \ref{caso1} we described the simple case when only 
        two images are being aligned. For more than two images, after
        calculating the canvas size ($W \times H$) and its central point
        ($\alpha_c , \delta_c$), the algorithm follows the next steps:

        \begin{enumerate}
        \item For each FITS image:
              \begin{itemize}
                \item Computing the pixel distance between center
		      of the $i$-th image and the coordinates ($\alpha_c$,$\delta_c$)
		      of the central pixel ($W/2$,$H/2$) of the canvas, by using
                      the equations shown in Section \ref{ecuaciones_desplazamiento}.
                \item Insert the image on the appropriate place.
                \item Make the rotation with Equation \ref{rot_equ}
                      for completing the alignment.

              \end{itemize}
        \item Write the data matrix on a  FITS file, 
              adding the correspondence between the central ($\alpha_c$,$\delta_c$) 
              coordinates and the central ($W/2$,$H/2$) pixel position in its
              header, and write the scale factor and other information in the header.

        \end{enumerate}

\subsection{Software details}

\label{apendice2}

\subsubsection{Pixel average}

	\label{nanproblem}

	The simple way for computing the average intensity of each pixel 
	(in MatLab\texttrademark) requires the use of the NaN 
        values for the surrounding  black area of the image. Actually, if GALEX
        images are processed with Montage, this NaN set up must be
        done by the user (with other software) or some strip effects could appear on the
        overlapping areas.

\begin{figure}[h!]
	\centering
        \advance\leftskip-1cm
	\includegraphics[trim=5mm 5mm -15mm 5mm,width=6cm,angle=270]{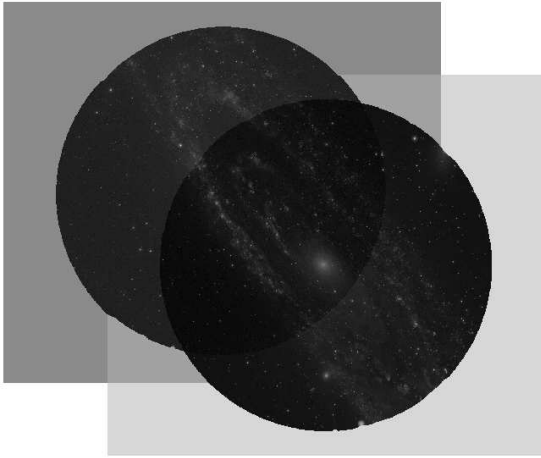}	
	\caption{Two merging FITS on one image per layer cube. Final average must be
          computed along the third dimension (perpendicular to the paper). 
          The NaN frame around the image is shown in gray 
          colors.}
        \label{averagex}
     \end{figure}

	A multi-layer matrix where these surrounding non-signal
	pixels have NaN values lets MatLab\texttrademark~ 
        use of \emph{nansum} function to
	add only data values. With this method, each layer will store
	only an image matrix at the first step of the process. When all the 
	images are placed in their correct position the average of the intensity
	is possible along the third dimension of the cube.
	Figure \ref{averagex} shows an example with two images correctly
	aligned. Each image is placed in its own layer. 
	Computing the resulting matrix in this case requires the before mentioned
	averaging.

	With the purpose of an effective use of the system resources (particularly
	the computer memory), a single-layer matrix can be used if the sum
	and average of each new incorporated image are made inside a bounded
	sub-matrix equivalent to the size of the new added image.

\subsubsection{Rotation}

	Image rotation requires some operations with pixel values that
	can create distortion because the presence of noise in the signal data.        
	The resulting matrix will have in each pixel a linear combination
	of original values. Our recommendation for completing the rotation
	process (in MatLab\texttrademark) involves the use of the
        \emph{imrotate} function applying the
        \emph{nearest neighbour} modality so
        that a loss-less interpolation be reached. This option searches, among 
        the original values, the most similar numbers to the estimated output.

     \begin{figure}[!ht]
	\centering
	\includegraphics[clip=true,trim=10mm 10mm 10mm 10mm,width=8cm]{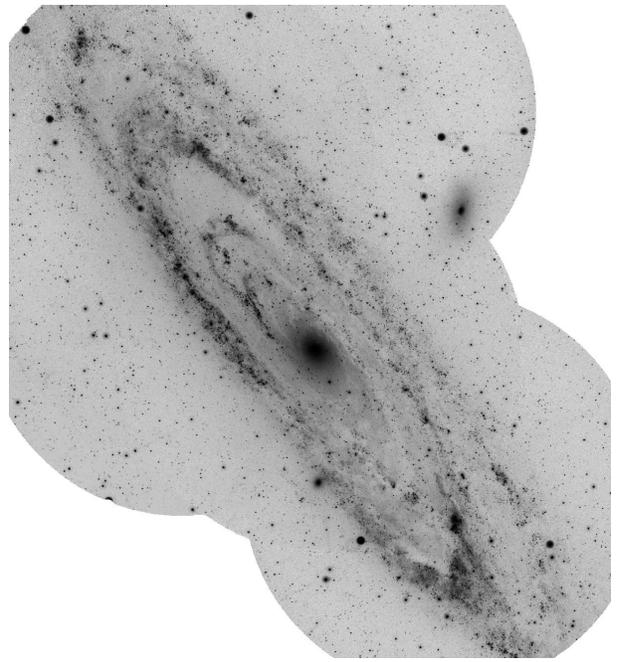}
	\caption{Final reconstruction of M31 galaxy from  several  FITS files.}
	\label{final}
     \end{figure}

\subsubsection{Availability}

\label{download}

MOSAIX is a set of MatLab\texttrademark ~scripts to build 
big mosaics from GALEX images. The original scripts are available at the 
\texttt{odin.estad.ucm.es/mosaix} web site. The scripts are prepared for a friendly 
use with three parameters: (1) the folder of the FITS files (2) the name of the output 
file and (3) a boolean flag for intermediate plots.
Get your mosaics typing in the MatLab\texttrademark ~prompt the next command:

~\begin{verbatim}
totalmerging('folder/','test.fits',0)
 \end{verbatim}

\section{Tests}

\label{test}

\subsection{Image quality}


The mosaicking process involves interpolation for rotation, averages 
of overlapping areas and numerical rounding. Comparing the result with the original 
data is the only way to 
check how much the data was modified, mainly when the aim of this task is
to do other kind of analysis with the final image. 

\begin{figure}[!ht]
	  \centering
	  \includegraphics[trim=20mm -5mm 20mm -40mm,width=5cm]{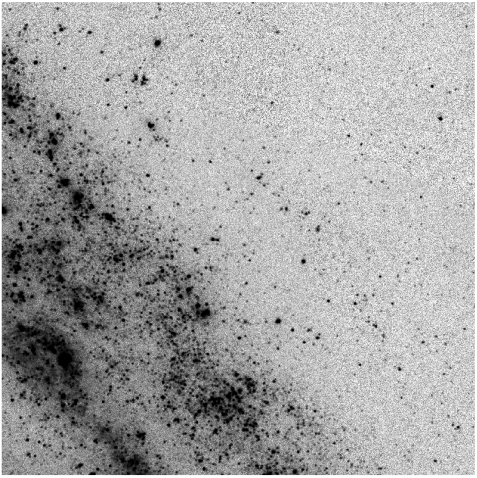}
	  \caption{Section of the final image where an overlapped area is included. Several subimages
          of this region have been processed in the image quality tests. The same section was taken 
          from MOSAIX as well as from Montage, for testing the similarity with the original image 
          (without overlapping).}
	  \label{M31section}
        \end{figure}

Seven FUV images from GALEX were used to generate our first mosaic, shown
in Figure~\ref{final}. For a reference comparison, the Montage tool was 
used too with the same set of images. Figure 
\ref{M31section} shows a $1000 \times 1000$ pixels section of the mosaic where
the quality tests were applied.

\begin{figure}[!ht]
	  \centering
	  \includegraphics[trim=-5mm -5mm 0mm 0mm,width=8cm]{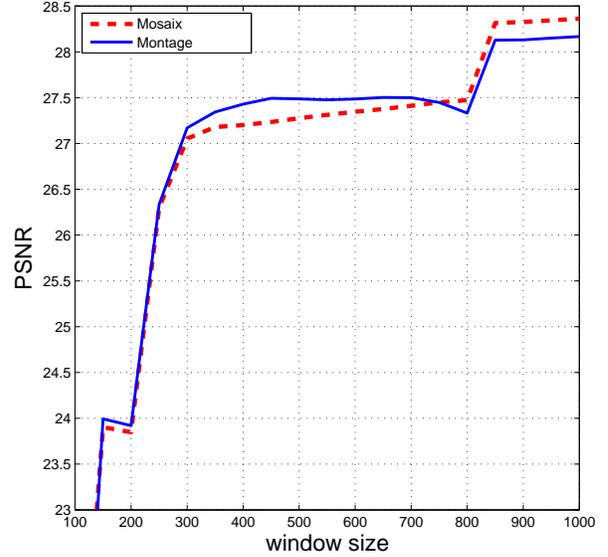}
          \includegraphics[width=8cm]{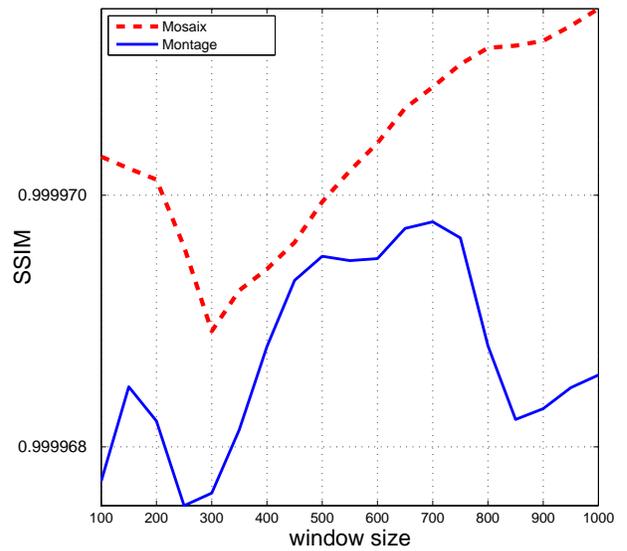}
	  \caption{Tests of comparison between original images and merged mosaics. On the top,
           PSNR measure is shown. On the bottom, the SSIM is
           shown in logarithmic axis. Both of them show the results for different window sizes of the image.
          The values are so close to the maximum value (1) that it was needed to zoom into the
          vertical axis.}
	  \label{psnrssim}
        \end{figure}

Two quality measures were used comparing the mosaic data with the original data. First,
the classical Peak Signal-to-Nosie Ratio (PSNR) defined as
$PSNR = 10 \log_{10} (MAX^2/MSE)$, where $MAX$ is the maximum possible value of the signal 
and $MSE$ is the mean square error between the original signal and the mosaic result.
Second, the Structural Similarity (SSIM) index which measures the similarity between 
two images and is defined as follows:
        \begin{eqnarray}
\hbox{SSIM}(x,y) = \frac{(2\mu_x\mu_y + c_1)(2\sigma_{xy} + c_2)}{(\mu_x^2 + \mu_y^2 + c_1)(\sigma_x^2 + \sigma_y^2 + c_2)}
          \end{eqnarray}
being $\mu_x$ the average of the first image (called $x$), $\mu_y$ the average of the second image (called $y$),
$\sigma_x^2$ the variance of $x$, $\sigma_y^2$ the variance of $y$ and $\sigma_{xy}$ the covariance of $x$ and $y$.
$c_1 = (k_1L)^2$ and $c_2 = (k_2L)^2$ are two variables to stabilize the division with weak denominator. 
The parameter $L$ is the dynamic range of the pixel-values, and $k_1 = 0.01$ and $k_2 = 0.03$ by default.
The performance of this measure has been shown recently by \cite{ssim}. Figure
\ref{psnrssim} shows the results for several image sizes. In both cases (SSIM and PSNR) the differences
are negligible, that is both programs produce final images that are very similar to the original
ones in the overlapped regions.

\subsection{Execution time}

\begin{figure}[!ht]
	  \centering
          \includegraphics[width=8cm]{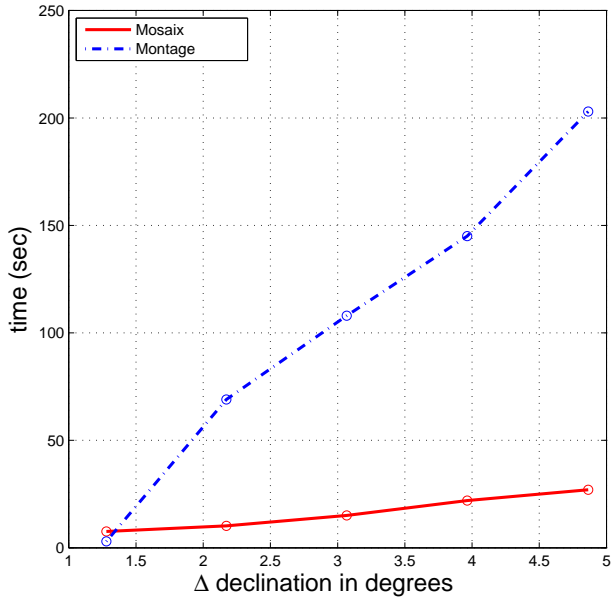}
	  \caption{Test of processing time. Several sized mosaics were processed
          with Montage software and our scripts. Vertical axis shows the time elapsed
          in each case.}
	  \label{time}
        \end{figure}

Montage software was developed to be able to work in several CPUs. However, 
for the common user the time to compute the total mosaic (in a single machine) is the 
most important factor to be considered. The timing tests were made on a multicore workstation 
with four cores for eight running threads Intel(R) Core\texttrademark i7 CPU 920 \@2.67 GHz and 6 Gb 
of RAM memory. Five overlapped images (increasing the mosaic field of view) were processed with the
Montage software. The same operations were 
processed with our scripts. To get the computing time the bash script \emph{time} command
was used for Montage and the \emph{cputime} MatLab\texttrademark ~function
for our scripts. Figure~\ref{time} shows that, as expected, the execution time increases 
with the size of the mosaic. However, our scripts show an improvement compared with Montage.

\section{Some conclusions and future work}

\label{epilogue}

This paper summarises the main characteristics of Mosaix, the tool developed by our 
team for mosaicking GALEX mission images. Our software accomplishes the standards of Montage 
in terms of overlapping quality and re-projection and guarantees no loss of information in 
the overlapping regions. As a specific tool for GALEX, our software is optimized 
to work with images from this mission. The user needs to provide only the original images from 
GALEX and run the programme. Previous operations with the GALEX 
images are not needed, contrarily to Montage. Velocity tests 
demonstrate our software is faster than Montage in single processor usage (we note that Montage 
can be run also in parallel mode). There are several improvements suitable to be incorporated in future 
versions such as the possibility of user-specified regridding or the generalization to images
from other missions. This initial version of our software is distributed
for MatLab\texttrademark. Subsequent versions for IDL will be
provided in the future.

\acknowledgments

	Thanks to Dr. Jason Dieter 
        Fiege\footnote{The \emph{mfits} library written by Dr. Jason Fiege 
          (\texttt{fiege@physics.umanitoba.ca}) at University of Manitoba is available for free.
        Some modifications were included in the scripts here developed.}, 
        writing resulting matrix in a  FITS file is 
        easy. With his \emph{mfits} library we can fit coordinates 
        and other header settings. We would like to thank the anonymous referee for
        many useful comments that improved this paper.
        We acknowledge financial support from Ministerio de Econom\'ia y Competitividad
        of Spain through grant AYA2011-29754-C03-01

\begingroup
\let\clearpage\relax

\bibliographystyle{spr-mp-nameyear-cnd}
\bibliography{armengot}

\endgroup

\end{document}